\documentclass[twocolumn,
reprint,
amsmath,amssymb,
aps, 
prl,
floatfix,
longbibliography]{revtex4-2}

\usepackage[FIGTOPCAP, hang]{subfigure}
\usepackage{float} 
\usepackage{algorithm}
\usepackage{algpseudocode}
\usepackage{algpseudocode}
\usepackage[english]{babel}

\usepackage{graphicx}
\usepackage{dcolumn}
\usepackage{bm}
\usepackage{hyperref}
\usepackage{lipsum}
\usepackage[super]{nth}
\usepackage{silence}
\usepackage{tikz-cd}
\usetikzlibrary{decorations.markings, arrows.meta}
\usepackage{esvect}
\usepackage{natbib}

\hypersetup{
colorlinks=true, 
citecolor=blue, 
urlcolor=blue, 
linkcolor=blue
}

\DeclareUnicodeCharacter{00A8}{}
\DeclareUnicodeCharacter{00B4}{}

\begin{document}

\setlength{\abovedisplayskip}{3pt}
\setlength{\belowdisplayskip}{3pt}

\preprint{APS/123-QED}

\title{Spectral Ranking of Causal Influence in Complex Systems}

\author{Errol Zalmijn$^{1,2}$, Tom Heskes$^1$, Tom Claassen$^1$}



\renewcommand{\andname}{\ignorespaces}
\affiliation{
 $^{1}$ Institute for Computing and Information Sciences, Radboud University, Nijmegen, the Netherlands
}
\affiliation{
 $^{2}$ ASML Research Department, Veldhoven, the Netherlands
}




\begin{abstract}
Like natural complex systems such as the Earth's climate or a living cell, semiconductor lithography systems are characterized by nonlinear dynamics across more than a dozen orders of magnitude in space and time. Thousands of sensors measure relevant process variables at appropriate sampling rates, to provide time series as primary sources for system diagnostics. However, high-dimensionality, non-linearity and non-stationarity of data remain a major challenge to effectively diagnose rare or new system issues by merely using model-based approaches. To reduce the causal search space, we validate an algorithm that applies transfer entropy to obtain a weighted directed graph from a system's multivariate time series and graph eigenvector centrality to identify the system's most influential parameters. The results suggest that this approach robustly identifies the true influential sources in a complex system, even when its information transfer network includes redundant edges.

\begin{description}
\item[PACS numbers]
\end{description}
\end{abstract}

\pacs{Valid PACS appear here}
\maketitle


Semiconductor lithography systems are extremely complicated electromechanical systems, capable of sub-nanometer positioning and sub-milliKelvin temperature control, while generating extreme ultraviolet light from laser-pulsed tin plasma.~It is notoriously difficult to fully understand the complex interactions among thousands of observed variables affecting the output of such systems.~Model-based approaches alone are inadequate to effectively diagnose rare or new system issues, as they inherently do not model abnormal behavior.~To efficiently, yet reliably reduce the search space of potential causes, we consider a system's causal network and the centrality of the network nodes. Therefore, we use Schreiber's transfer entropy \cite{tschreiber2000} which quantifies predictive information transfer or influence between two time series, providing direction, strength and delay of both linear and non-linear interactions. For time series with a Gaussian distribution, transfer entropy reduces to Granger causality \cite{lbarnett2009}. However, being a bivariate measure, transfer entropy naturally disregards the multivariate nature of interactions in complex systems. An exhaustive multivariate decomposition into redundant, unique and synergistic information would become computationally intractable, due to combinatorial scaling with the number of involved variables. We mitigate this curse of dimensionality, by combining transfer entropy as measure of a node's direct influence on its neighbors with eigenvector centrality \cite{pbonacich1972} as measure of a node's global influence in the network, to identify the system's most important propagation sources of original information or influence.\\
\indent In this paper, we introduce the measures of transfer entropy and eigenvector centrality and describe the two-step algorithm as used.~In rolling window analysis of a simulated time series representing two bidirectionally coupled Lorenz systems, we compare its causal inference results to those of a constraint-based algorithm for multivariate causal analysis.~The rolling window approach allows additional analysis of time-evolving global influence of the Lorenz system state variables. Finally, we assess the algorithm in diagnosing a real world industrial system issue, using high-dimensional time series.\\
\indent Transfer entropy ($TE$) is an information-theoretic implementation of Wiener's notion of causality applied to time series \cite{nwiener1956}, whereby the cause precedes - and contains unique information about - the effect. Consider two stationary ergodic Markov processes $X$ and $Y$ (or $X^{(i)}$ and $X^{(j)}$ as below) and their corresponding time series ${\lbrace x_1, x_2, ..., x_M \rbrace}$ and ${\lbrace y_1, y_2, ..., y_M \rbrace}$ of $M$ samples. Transfer entropy quantifies reduction in uncertainty about future states of a source process $X$, when passed states of a target process $Y$ are observed in addition to passed states of $X$ itself. As an asymmetric measure based on transition probabilities, transfer entropy naturally incorporates directional and dynamic information which may imply causation between $X$ and $Y$:

\onecolumngrid

\begin{align}
TE_{X\rightarrow Y}^{(k,l)}(t, \tau) &= \displaystyle \sum_{y_{t},y^{(l)}_{t-1}, x^{(k)}_{t-\tau}} p(y_t,y^{(l)}_{t-1}, x^{(k)}_{t-\tau})\log_{b}\frac{p(y_t|y^{(l)}_{t-1}, x^{(k)}_{t-\tau})}{p(y_t|y^{(l)}_{t-1})}\label{eq1}
\end{align}

\twocolumngrid


\noindent where $x_t$ and $y_t$ represent states of $X$ and $Y$ at time $t$ while \small $TE_{X\rightarrow Y}^{(k,l)}(t, \tau)$ \normalsize indicates maximized information transfer from $X$ to $Y$, computed across a range \small ${D=\{0,1,2,...,\tau_{max}\}}$ \normalsize of embedding delays $\tau$, such that \small $\displaystyle \max_{\tau \epsilon D} \{TE_{X\rightarrow Y}^{(k,l)}(t, \tau)\} = TE_{X\rightarrow Y}^{(k,l)}(t,\ \displaystyle \operatorname*{arg\,max}_{\tau \epsilon D} \{TE_{X\rightarrow Y}^{(k,l)}(t,\tau)\})$. \normalsize The embedding dimensions $k$ and $l$ denote the number of passed states in $X$ and $Y$ used to condition the probabilities of transition to the next state of $Y$ (or $X$) represented by ${x^{(k)}_{t-\tau}=\lbrace x_{t-\tau-k+1},x_{t-\tau-k+2},...,x_{t-\tau}\rbrace}$ and ${y^{(l)}_{t-1}=\lbrace y_{t-1-l+1},y_{t-1-l+2},...,y_{t-1}\rbrace}$.~The logarithm \mbox{base $b=2$}, defines the informational unit of transfer entropy in bits i.e.\ reduction in average code-length required to optimally encode the target variable (effect), given passed states of the source variable (cause) and target variable.~Herein, we keep the embedding dimensions at the commonly used ${k = l = 1}$, mostly for computational reasons. For every pair ${(X^{(i)}, X^{(j)})}$ from $N$ multivariate time series ${X^{(1)}, X^{(2)}, ... , X^{(N)}}$, we apply Eq.~(\ref{eq1}) to estimate information transfer $TE$ and thereby determine the candidate source times series $X_{(d)}$ and target time series $X_{(r)}$.~We then generate ${1/ (\alpha=0.05) -1}$ surrogates ${\{X^{'(1)}_{(d)},X^{'(2)}_{(d)},...,X^{'(19)}_{(d)}\}}$ which share their amplitude distribution and power spectrum with original time series $X_{(d)}$, using the iterative amplitude adjusted Fourier transform (iAAFT) proposed by Schreiber et al.~\cite{iaaft}.~Following this study, we estimate information transfer $TE^{'}$ from each surrogate to (original) target time series $X_{(r)}$ and obtain ${\{TE^{'}_{1},...,TE^{'}_{19}\}}$.~If ${TE > max \{TE^{'}_{1},...,TE^{'}_{19}\}}$, information transfer $TE$ is considered to be significant or non-significant otherwise. The resultant information transfer network, given by a directed weighted graph ${G = (V,E)}$, comprises a set ${V = \{v_i\}_{i=1}^{N}}$ of $N$ nodes and \mbox{$E = \{e_{ij} =\ (v_i, v_j)\}$} of edges. Each edge $e_{ij}$ connects a source node $v_i$ to target node $v_j$ with strength $w_{ij}$ i.e. information transfer $TE$.\\
\indent To diagnose performance issues or even failures within technological complex systems, we wish to locate where perturbations enter the system and propagate, causing downstream effects throughout the system.~Therefore, we consider an information transfer network wherein the (main) sources of original information can be identified by measurement and ranking of each node's global network influence termed centrality or importance.~Centrality, the basic principle of Google's search engine \cite{lpage1998}, has proven useful to measure and rank a page's relevance based on its inbound links. Here, we use out-degree eigenvector centrality which defines the centrality $c(v_i)$ of a node \mbox{$v_i\ \epsilon\ V= \{v_1, ... ,v_n\}$} as proportional to the summed centralities of its outbound neighbors:
\begin{align}
c(v_i) = \displaystyle \lambda^{-1} \sum_{j=1}^N W_{ij} c(v_j) & \text{, or } Wc =\lambda c \label{eq3}  
\end{align}
\noindent
 where $\lambda^{-1}$ is a proportionality factor and $c$ is the eigenvector of centralities associated with eigenvalue $\lambda$ of adjacency matrix $W$, whose entries $w_{ij}$ denote information transfer from node $v_i$ to node $v_j$ in graph $G$. If $W$ is non-negative and irreducible, the Perron-Frobenius theorem \cite{gfrobenius1912, operron1907} ensures there is a unique vector $c_{1}$ of $N$ centralities $c_1(v_i) > 0\ {,} \forall\ v_i$ associated with the largest positive eigenvalue ${\lambda_{1} = \rho(W)}$ or spectral radius of $W$, satisfying Eqs.~(\ref{eq3}).~Usually $c_1$ is normalized, such that each entry indicates the centrality or importance of a node $v_i$ in graph $G$ on a relative scale from 0 to 1.~Alternatively, matrix $W$ is normalized to a transition matrix $P$, whose entries $p_{ij}$ denote probabilities of transition from node $v_i$ to node $v_j$ in a random walk on graph $G$ or Markov chain while $\sum_{j=1}^{N} p_{ij} = 1$, such that:   


\begin{align}
p_{ij}= \begin{cases} w_{ij}/\sum_{j=1}^{N} w_{ij},  & \mbox{~~if~~} \sum_{j=1}^{N} w_{ij} \neq 0\\[4pt] 0, & \hspace{0,75cm} otherwise
\end{cases}
\end{align}

\noindent Following Google’s PageRank approach \cite{lpage2006}, we modify matrix $P$ by adding a teleportation probability $\gamma\ \epsilon\ \langle 0,1\rangle$ and an all-ones matrix $J$ to obtain an ${N \times N}$, irreducible, positive matrix $P^{'}$: 
\begin{align}
P^{'}_{\gamma} = \gamma P + \frac{1}{N}(1 - \gamma)J\label{eq5}
\end{align}
where a random walker at node $v_i$ follows an edge with probability $\gamma$ or jumps to any other node in the $N$-nodes network with probability ${(1-\gamma)}$. Herein, we use PageRank's typical value of ${\gamma = 0.85}$. Considering the Markov chain associated with matrix $P^{'}_{\gamma}$, the Perron-Frobenius theorem ensures an unique stationary probability distribution that matches the eigenvector ${\vec{\pi}_{1}}$ of ${P^{'}_{\gamma}}$ associated with eigenvalue ${\lambda_{1} (P^{'}_{\gamma}) = 1}$, such that $\vec{\pi}_{1} = P^{'}_{\gamma} \vec{\pi}_{1}$. Eigenvector $\vec{\pi}_{1}$ is usually computed via power iteration \cite{pwr_itr_v_mises} in $k$ steps: ${\vec{\pi}_{1} = \underset{k \to \infty}{\lim} \vec{\pi}^{(k)} = \underset{k \to \infty}{\lim} P^{'(k)}_{\gamma} \vec{\pi}^{(0)}}$ where ${\vec{\pi}^{(0)}}$ denotes an initial distribution. Empirical studies by Kalavri et al.~\cite{kalavri_v_2016} revealed that PageRank (or eigenvector centrality) yields similar ranking results when computed with, or without a graph's (first-order) semi-metric edges, which we consider sufficient to robustly and reproducibly identify a system's key propagation sources of influence.\\ 
\indent In what follows, we assess the method's accuracy in causal inference and node ranking using simulated time series, followed by root cause analysis from real world diagnostic data.~Firstly, we consider a system ${S_{(L_1 \rightleftarrows L_2)}}$ of two bidirectionally coupled Lorenz systems $L_1$ and $L_2$, investigated by Wibral et al.~\cite{mwibral2012} and given by:


\small
\begin{subequations}
\renewcommand{\theequation}{\theparentequation.\arabic{equation}}
\begin{align}
\dot{X}_1(t) &=10.0(Y_1(t)-X_1(t))\label{eq51}\\   
\dot{Y}_1(t) &=X_1(t)(25.0-Z_1(t))-Y_1(t)+0.1Y_2^2(t-3)\label{eq52}\\
\dot{Z}_1(t) &=X_1(t)Y_1(t)-2.67Z_1(t)\label{eq53}\\ 
\dot{X}_2(t) &=10.0(Y_2(t)-X_2(t))\label{eq54}\\ 
\dot{Y}_2(t) &=X_2(t)(28.0-Z_2(t))-Y_2(t)+0.05Y_1^2(t-5)\label{eq55}\\
\dot{Z}_2(t) &=X_2(t)Y_2(t)-2.67Z_2(t)\label{eq56}
\end{align}
\end{subequations}

\normalsize
\noindent
 The bidirectional coupling ${Y_1 \rightleftarrows Y_2}$ is governed by time-delayed quadratic terms.~We used \mbox{Pydelay \cite{vflunkert2011}} to generate a multivariate time series \small \mbox{$\{X_1, X_2,Y_1,Y_2,Z_1,Z_2\}$} \normalsize of 150K samples by numerically integrating \mbox{Eqs. (\ref{eq51} - \ref{eq56})} with step size ${dt=0.01}$ and initial conditions \small \mbox~{$X_1(0)=X_2(0)=1.0$, $Y_1(0)=Y_2(0)=0.97$} \normalsize and \small \mbox{$Z_1(0)=Z_2(0)= 0.99$}. \normalsize We use the FaultMap algorithm~\cite{sstreicher2014} which, to our knowledge, is the only publicly available implementation of the described approach. To assess its accuracy in causal network reconstruction against other model-free approaches, we use PCMCI (v4.0) \cite{jrunge2018} as a distinct, constraint-based, multivariate alternative.~FaultMap estimates edge-weight $w_{ij}$ as ${\Delta TE(\tau) = TE_{X^{(i)} \rightarrow X^{(j)}}(\tau) - TE_{X^{(j)} \rightarrow X^{(i)}}(\tau)}$ using the Kraskov-Stögbauer-Grassberger estimator for transfer entropy in the Java Information Dynamics Toolkit \cite{jidt}. PCMCI performs condition-selection at an optimized significance level $\alpha$ using Akaike's information criterion, followed by a conditional independence test where we use the linear ParCorr test (at ${\alpha=0.05}$) instead of the computationally expensive nonlinear CMI test. Following the recommended sample size of Bauer \cite{m_bauer2005}, we use a rolling analysis window $W_{s}$ of 2K samples to define 50 adjacent time series slices for causal network reconstruction of the coupled Lorenz systems by both algorithms and node importance ranking by FaultMap only. Figure 1 depicts the rolling window analysis approach, the butterfly-shaped attractors of the coupled Lorenz systems in phase space $(x,y,z)$ and cause-effect detection count heatmaps for both algorithms. Figure 2 exemplifies FaultMap's rolling window analysis results of which Figure 3 specifically reports information transfer (delay) via coupling ${Y_1 \rightleftarrows Y_2}$ and global influence of $X$-, $Y$- and $Z$-variables on the coupled Lorenz systems. The heatmaps in Figure 1b show the total count of each cause-effect relation detected by FaultMap or PCMCI over 50 adjacent time series slices. FaultMap detected the bidirectional coupling ${Y_1 \rightleftarrows Y_2}$ at a detection rate of 89 \% vs 85 \% by PCMCI, while the latter achieved a notably higher overall detection rate (95 \%) of causal links within the Lorenz subsystems than FaultMap \mbox{(83 \%)}. This difference is mainly related to PCMCI's apparent higher sensitivity to autocorrelative effects encoded in the time series. \noindent
PCMCI also detected 23 \% more transitive indirect links than FaultMap, while we would expect mere direct relations from its multivariate causal analysis.
\raggedbottom
 
\begin{figure}[htp]
\centering
\subcapnoonelinetrue
\subfigure[\ Lorenz time series, rolling analysis window and attractors.]
{\includegraphics[width=\columnwidth]{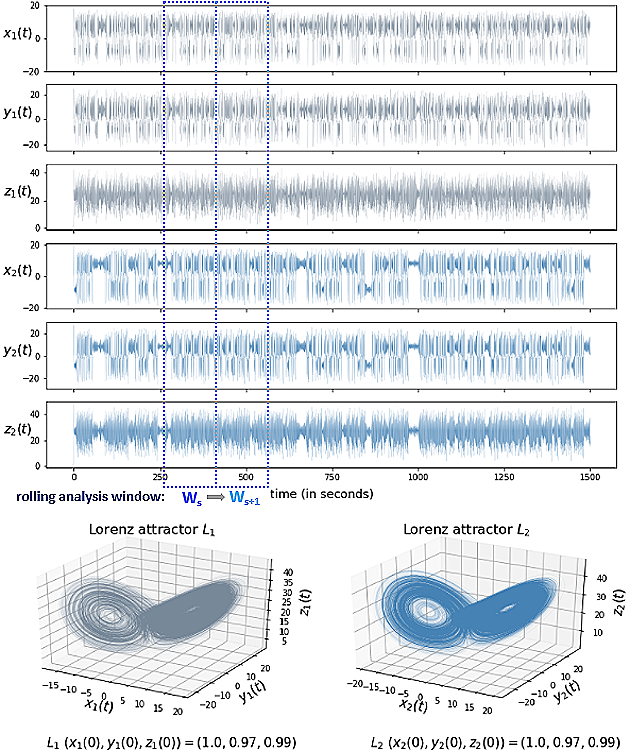}}
\subfigure[\ cause-effect detection count over 50 adjacent time series slices]{\includegraphics[width=\columnwidth]{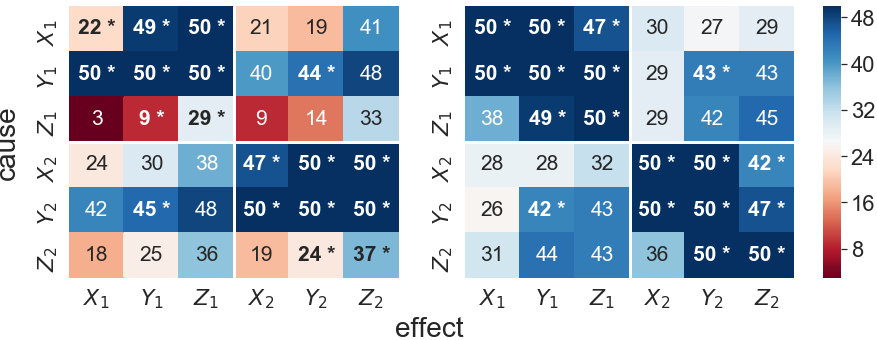}}
\caption{(a) time series of system ${S_{(L_1 \rightleftarrows L_2)}}$ composed of bidirectionally delay-coupled Lorenz systems $L_1$ and $L_2$, generated from Eqs. (\ref{eq51} - \ref{eq56}). (b) heatmaps of detection count per cause-effect relation, for FaultMap (left) and PCMCI (right). Direct cause-effect relations are denoted by (*).}
\end{figure} 

\indent Due to transitivity of bivariate information transfer, most networks inferred by FaultMap include direct and indirect connections, all of which passed strict significance tests, as \mbox{$Y_1 \rightarrow Y_2 \rightarrow Z_2$} and \mbox{${Y_1 \dashrightarrow Z_2}$} shown in Figure 2.~The coupled Lorenz systems are easily discernible by two subnetworks \mbox{$(X_1,Y_1,Z_1)$} and \mbox{$(X_2,Y_2,Z_2)$} that exhibit distinct levels of information transfer with similar time delays in the range of milliseconds. At a rate of at least 98 \%, FaultMap detected the same edges \mbox{$X \rightarrow Y$}, \mbox{$Y \rightarrow X$}, \mbox{$X \rightarrow Z$} and \mbox{$Y \rightarrow Z$} indicating (normalized) net influence within both Lorenz subsystems, as Gencaga et al. \cite{Gencaga2015} found for a single Lorenz system.~PCMCI shows similar detection results for these edges (see heatmaps).~Of note, the rolling window approach took PCMCI 495 min of runtime on a 24-cores HPC node with 192 Gb for causal analysis vs 58 min by FaultMap for causal analysis and node ranking.\\
\indent In Figure 3a-b, we focus on reconstruction of time delays in coupling ${Y1 \rightleftarrows Y2}$, given the modeled delays in \mbox{Eq. (\ref{eq52})} and (\ref{eq55}). Figure 3a reveals the dynamic nature of bidirectional information transfer in terms of strength and delay. The median difference ($\approx$ 0.05) of information transfer distributions in Figure 3b suggests that Lorenz system $L_2$ predominantly drives $L_1$, which is reasonable to expect since the coupling strength ($0.1$) in \mbox{Eq. (\ref{eq52})} is twice the coupling strength ($0.05$) in \mbox{Eq. (\ref{eq55})}.~Given coupling delays of 3 and 5 sec, the distribution of reconstructed interaction delays seems realistic but may be impacted by lag synchronisation of the Lorenz systems, as suggested by Coufal et al.~\cite{coufal_d_2017}. Figure 3c shows highly dynamic global influence of particularly $X$- and $Y$-variables, while the $Y$-variables remain the driving force within their respective Lorenz subsystem at all times.~Grey-colored bars highlight time windows in which $L_1$ is identified as driving, and $L_2$ as driven subsystem.~Since the coupling strengths are constant, ${Y_1 \rightarrow Y_2}$ is likely to dominate ${Y_2 \rightarrow Y_1}$ in strength when the $Y_2$ state reaches vanishingly low values relative to the $Y_1$ state.~The median difference across all node importance distributions in Figure 3d reflects the aforementioned drive-response relations in, and between the interacting Lorenz systems.~It might explain FaultMap's 100 \% detection rate of $Y$-variable self-loops vs lower rates of all other self-loops.~The outcome of both $Y$-variables as Lorenz system key driver complies with the Lorenz model of Rayleigh-Bénard convection~\cite{rayleigh_l_1916}, where temperature difference drives convective heat transfer in addition to conduction at Rayleigh number ${Ra \geq 25}$ (see \mbox{Eqs. (\ref{eq52}) and (\ref{eq55}))}. To our knowledge, this is the first rolling window analysis to date, capturing the dynamics of time-varying information transfer or global influence (importance) of state variables within interacting Lorenz systems.~Our findings may enable automated identification of monitoring observables for performance (anomaly) diagnostics or predictive maintenance within technological complex systems.~The ability to capture time-varying importance of a complex system's state variables is also relevant in time series analysis of natural complex systems, including the Earth's climate.\\

\begin{figure}[htp]
\centering
\includegraphics[width=0.95\columnwidth]{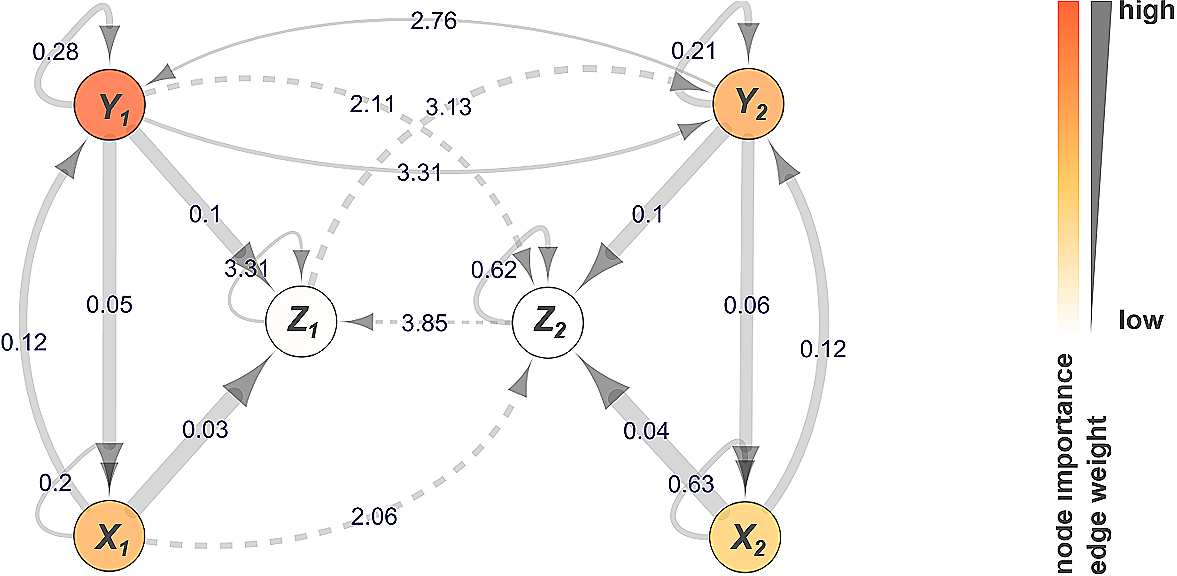}
\caption{Information transfer network of two bidirectionally delay-coupled Lorenz systems.~Edges indicate direct ($\longrightarrow$) or transitive indirect ($\dashrightarrow$) information transfer.~Edge annotations denote information transfer delay (sec).~Node importance indicates a node's global network influence.~Edge-weight represents level of information transfer ($w_{ij}$).}
\end{figure}

\begin{figure}[htp]
\centering
\subcapnoonelinetrue
\subfigure[dynamic information transfer (delay) via bidirectional delay-coupling ${Y_1 \rightleftarrows Y_2}$ between Lorenz systems $L_{1}$ and $L_{2}$]{\includegraphics[width=\columnwidth]{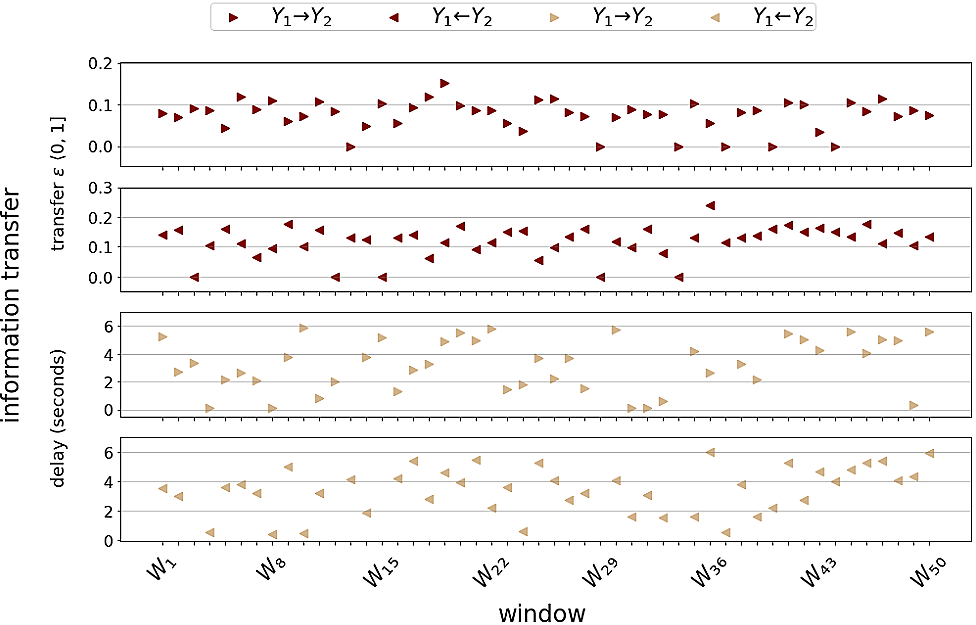}}
\subfigure[\ distribution of information transfer (delay)  in 3a  \hspace{0.5cm}]{\includegraphics[width=\columnwidth]{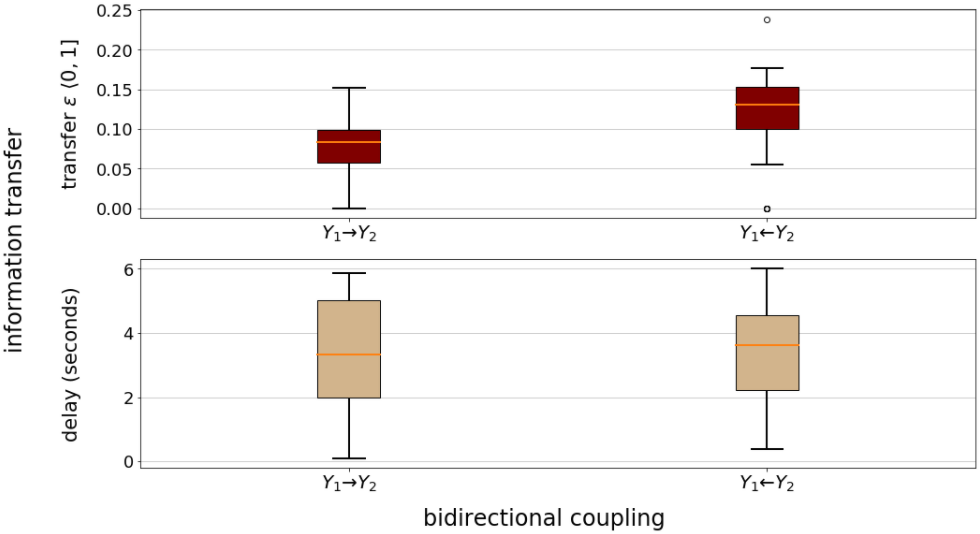}}
\subfigure[\ dynamic importance of state variables in delay-coupled Lorenz systems $L_{1}$ and $L_{2}$ \hspace{0.5cm}]{\includegraphics[width=\columnwidth]{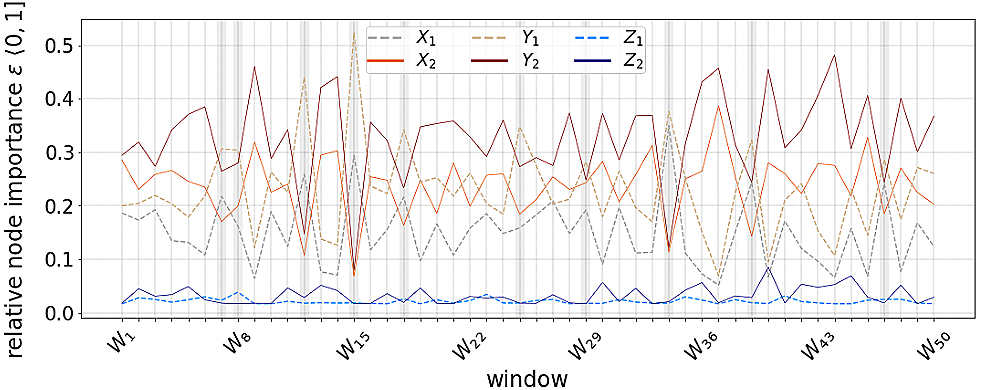}}
\subfigure[\ distributions of Lorenz system state variable importance in 3c \hspace{0.5cm}]{\includegraphics[width=\columnwidth]{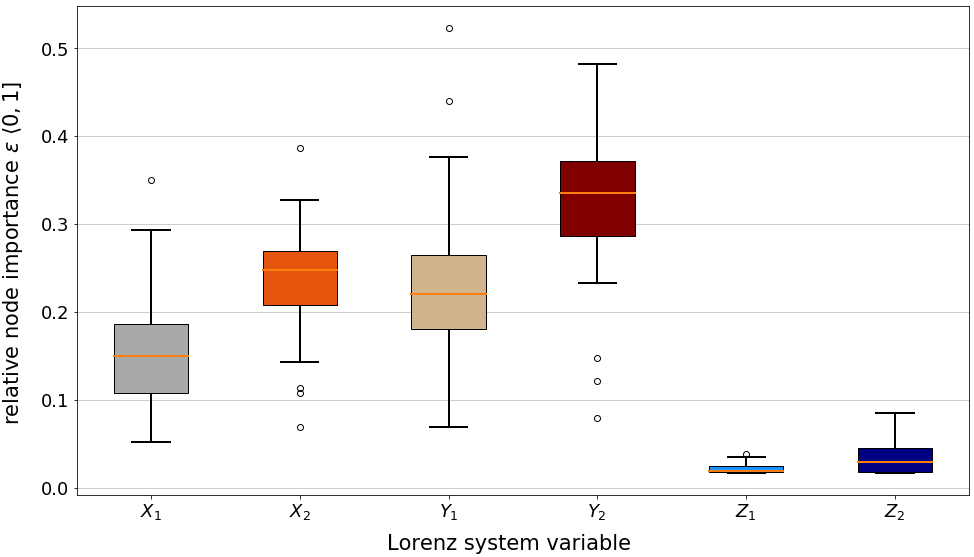}}
\caption{dynamic information transfer via bidirectional delay-coupling ${Y_1 \rightleftarrows Y_2}$ of Lorenz systems $L_{1}$ and $L_{2}$, and dynamic importance of the Lorenz system state variables.}
\end{figure}

\indent Nanolithography systems are among the most complex technological systems today, capable of sub-nanometer positioning and sub-milliKelvin temperature control, even as system modules accelerate at up to $15Gs$.~Such systems are particularly challenging for model-based diagnosis of rare or new issues, due to nonlinear interactions across multiple time and spatial scales.~To assess FaultMap's potential in diagnosis of issues within such systems, we investigate temperature, flow and pressure instability within an ASML subsystem.~Therefore, we use a multivariate time series of $315$ binary samples from $366$ parameters related to the problem.~As shown in Figure 4, FaultMap identified parameter $P_{0}$ as primary source of original information i.e.\ most probable cause leading to event $P_{27}$, through a network of collateral effects $\{P_1, ... ,P_{26}\}$.~The indicated root cause is confirmed to be correct by a series of automatically logged system events as well as service actions. This preliminary result is a promising step in automation of reliable data-driven diagnostics for technological complex systems.

\begin{figure}[htp]
\centering
\subfigure{\includegraphics[width=\columnwidth]{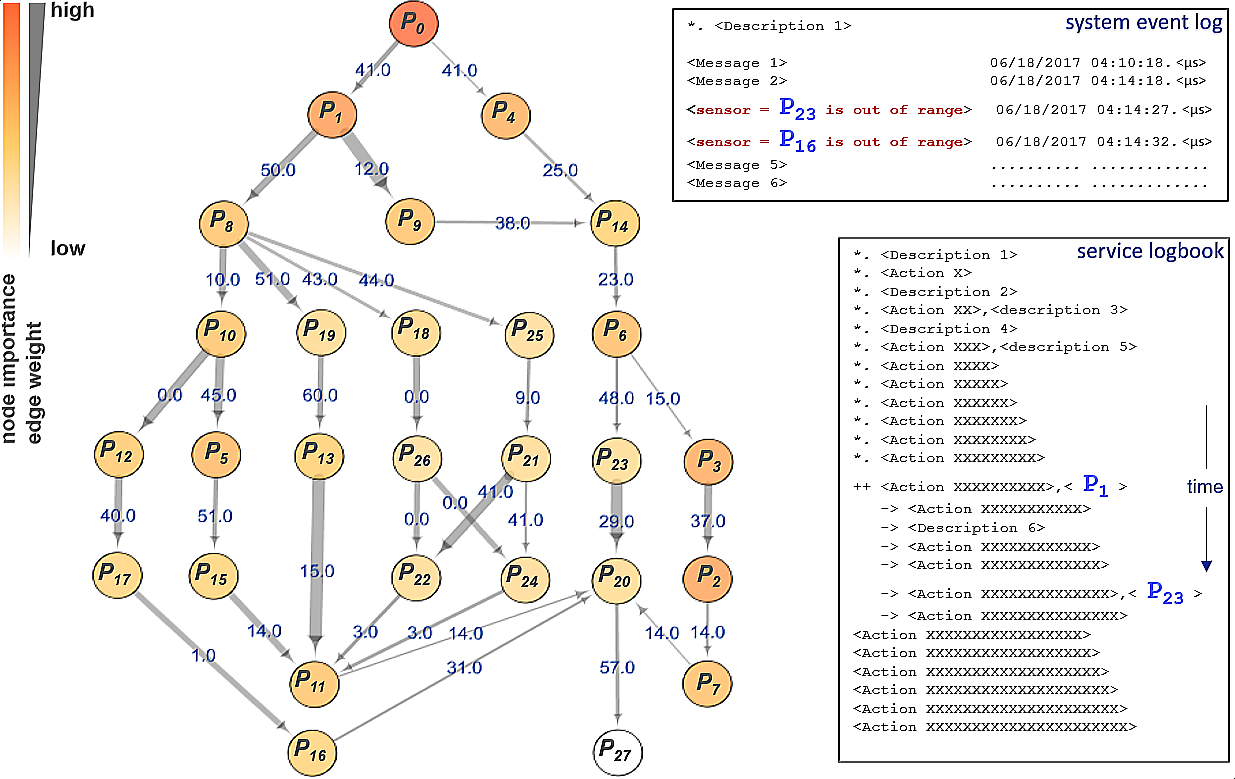}}
\caption{Top-ranked node $P_{0}$ (root cause) transfers original information towards event $P_{27}$ via a network of collateral effects $\{P_1, ... ,P_{26}\}$ within an ASML subsystem. The legend of Figure 2 applies here.}
\end{figure}

 \indent To fully understand a complex system's (deviant) behaviour, it is essential to identify its main propagation sources of influence affecting downstream elements throughout the system. We empirically show that spectral centrality analysis of the related information transfer network allows to robustly and reproducibly identify a complex system's main sources of original information or influence.~Our only algorithm of choice compares favorably against the sufficiently distinct alternative algorithm in causal analysis of two nonlinearly coupled Lorenz systems.~Also, it shows to be accurate, robust and efficient, identifying the (alternately) driving and driven Lorenz subsystems as well as the driving force within either subsystem. Finally, the algorithm correctly locates the original disturbance within a technological complex system. We conclude that the inherent robustness of spectral centrality ranking to graph semi-metricity, allows to reliably and efficiently identify a complex system's most important propagation sources of influence, without the need to differentiate between direct and indirect links.~This finding is relevant for any time series analysis of natural and artificial complex systems.\\
 \indent We thank David Sigtermans for our many inspiring discussions and Leonardo Barbini for his valuable feedback on the causal analysis results of the Lorenz systems.~Finally, we are grateful to Simon Streicher for his implementation and  helpful suggestions using it.

\bibliography{\jobname}
\end{document}